\documentstyle[12pt]{article}
\begin{document}
\begin{titlepage}

\hfill{UM-P-93/51}

\hfill{OZ-93/11}

\hfill{hep-ph/9308256}

\vskip 1cm

\centerline{\Large \bf Study of the simplest realistic Higgs sector}
\vskip 2 mm
\centerline{\Large \bf in quark-lepton symmetric models}

\vskip 1.5cm

\centerline{{\large Y. Levin}\footnote{email:
yal@tauon.ph.unimelb.edu.au}
{\large and R. R. Volkas}\footnote{email:
U6409503@hermes.ucs.unimelb.edu.au}}

\vskip 1.5cm

\noindent
\centerline{\it Research Centre for
High Energy Physics, School of Physics,}
\centerline{\it University of Melbourne, Parkville 3052, Australia}

\vskip 1.5 cm

\centerline{ABSTRACT}

\noindent
A discrete symmetry between quarks and (generalized)
leptons can exist in nature, and
its spontaneous symmetry breaking scale can be as low as a few TeV.
Such a discrete symmetry also has interesting implications for how
electroweak symmetry is spontaneously broken, because the simplest
version of the theory requires two electroweak Higgs doublets rather
than one in order to provide acceptable values for quark and lepton
masses. The effective theory generated at electroweak-scale energies is
thus a particular type of two-Higgs-doublet model. We point out in this
paper that the broken discrete symmetry imposes very interesting
constraints on the form of the Yukawa couplings between physical Higgs
bosons and quarks and leptons. In particular, we find that the
flavour-changing neutral Higgs couplings to down-sector quarks are
proportional to the neutrino Dirac mass matrix. If neutrinos are Dirac
particles, then the severe experimental upper bounds on their mass values
renders tree-level neutral flavour-changing Higgs effects on down-quark
systems like $K^0-\bar{K}^0$ negligibly small. We also discuss
minimization of some relevant Higgs potentials and some other pertinent
phenomenological issues.
\end{titlepage}

\leftline{\large \bf 1. Introduction}
\vskip 5mm

If there is physics beyond the Standard Model (SM), then it probably
involves at least one new symmetry principle of nature. Our experience
with the interactions of quarks and leptons strongly suggests that the
search for new symmetries is likely to bear fruit, because symmetries
play a central role in the SM.

If we adopt a ``bottom-up'' approach to model-building -- that is, if
our starting point is what we know of low-energy particle interactions
rather than an ambitious unifying principle of some sort -- then the
first new invariance we might hope to
uncover at some energy scale $> 100$ GeV is likely
to be a discrete symmetry. This is a reasonable suggestion simply
because discrete symmetries are the simplest candidates. For instance,
one may like to suppose that the complete Lagrangian of the world,
describing some fundamental unified theory, displays some large,
elegant, continuous
invariance group which is broken in many stages down to $G_{SM} = $
SU(3)$_c \otimes$SU(2)$_L \otimes$U(1)$_Y$ and finally just SU(3)$_c
\otimes$U(1)$_Q$. It could well be that the first enlargement of the
symmetry group of nature
above the electroweak scale involves some discrete symmetry subgroup of
the large fundamental invariance group. Discrete subgroups might well
survive intact down to quite low energies because they yield less new
physics than either global or local continuous symmetries, and are thus
likely to be less phenomenologically constrained. Alternatively, it may
turn out that discrete symmetries are of greater fundamental importance
than current theoretical prejudices allow.

If we look at a quark-lepton generation,
\begin{eqnarray}
& Q_L \sim (3,2)(1/3),\quad u_R \sim (3,1)(4/3),\quad d_R \sim
(3,1)(-2/3) &\ \nonumber\\
& \ell_L \sim (1,2)(-1),\quad e_R \sim (1,1)(-2),\quad \big[\nu_R \sim
(1,1)(0)\big], &\
\end{eqnarray}
where the quantum numbers are given with respect to $G_{SM}$, then three
generic classes of discrete symmetries suggest themselves: (i)
horizontal, (ii) left-right and (iii) quark-lepton symmetries.
Horizontal symmetries are the simplest to implement in the sense that no
extension to the gauge group $G_{SM}$ is mandatory. Left-right symmetry
(either parity or charge-conjugation invariance) can be implemented if
we extend $G_{SM}$ to the left-right group $G_{LR} = $SU(3)$_c
\otimes$SU(2)$_L \otimes$SU(2)$_R \otimes$U(1)$_{B-L}$. This gauge group
extension requires the inclusion of a new fermion -- the right-handed
neutrino -- whose presence is optional in the SM.
Quark-lepton
discrete symmetry (q-$\ell$ symmetry for short) can be implemented if we
extend $G_{SM}$ to the new group $G_{q\ell}$ where
\begin{equation}
G_{q\ell} = {\rm SU(3)}_{\ell} \otimes {\rm SU(3)}_q \otimes {\rm
SU(2)}_L \otimes {\rm U(1)}_X,
\label{Gqell}
\end{equation}
where SU(3)$_{\ell}$ is a ``leptonic colour'' group and SU(3)$_q$ is
just the usual colour group with a new name \cite{FL-1}.
This gauge group extension
also requires the introduction of new fermions, in this case the
leptonic colour partners of standard leptons (as well as a right-handed
neutrino).

Horizontal symmetries, discrete or otherwise, and left-right symmetry
have been assiduously studied for the past twenty years or so, and they
remain very important and interesting possibilities for new physics. The
possible existence of a quark-lepton discrete symmetry has, however,
only been pointed out quite recently, and so much work remains to be
done in this area. Actually, some work has already been
performed on neutral current
phenomenology \cite{FLV-2}, partial unification schemes \cite{FLV-3},
the fermion mass problem \cite{FL-4} and
cosmological implications \cite{LV-5,FLV-6}.
However, two important aspects of q-$\ell$
symmetric models require more attention in the literature. The first
area concerns the phenomenology of the new strongly-interacting sector
predicted by the theory. [An SU(2) subgroup of leptonic colour
remains unbroken and confines the exotic partners of the standard
leptons into unstable, non-relativistic bound states.] Although some
initial studies were conducted in Refs.~\cite{FLV-6,glueballs-7},
much more detailed work is
required. The other area is the subject of this paper: the phenomenology
of the extended Higgs sector of q-$\ell$ symmetric models.

Quark-lepton symmetric models can employ a number of different types of
Higgs sectors, depending on what one wants to do exactly. For instance,
if one wishes to employ the see-saw mechanism \cite{seesaw-8}
for neutrino masses then a
more complicated Higgs sector is required than if one just fine-tunes small
neutrino masses. Also, the q-$\ell$ discrete symmetry can induce
troublesome mass relations between quarks and leptons if the Higgs
sector is too simple. It is possible to adopt different attitudes to
what one should do about this problem, and this leads to different Higgs
physics. There is no clearly preferred option for the Higgs sector at
the moment. In this paper we will therefore review the major possibilities,
but will ultimately concentrate on the detailed phenomenology of a particular
concrete scenario for reasons we will make clear later.

The rest of this paper is structured as follows: In Sec.~II we review
the possible choices for Higgs sectors in q-$\ell$ symmetric models.
Section III forms the core of our paper. We
study one simple and workable scenario in
detail. We look at (i) the construction and minimization of Higgs
potentials, (ii) the identification of the physical Higgs fields and
their Yukawa couplings, and (iii) the derivation of bounds from
tree-level flavour-changing effects induced by the neutral Higgs bosons.
Our concluding discussion forms Sec.~IV.

\vskip 1 cm
\leftline{\large \bf 2. Higgs sectors for quark-lepton symmetric models.}
\vskip 5 mm

The simplest gauge group which supports discrete q-$\ell$ symmetry is
given by $G_{q\ell}$ in Eq.~(\ref{Gqell}). A fermionic generation is given
by
\begin{eqnarray}
& Q_L \sim (1,3,2)(1/3),\quad u_R \sim (1,3,1)(4/3),\quad d_R \sim
(1,3,1)(-2/3), &\ \nonumber\\
& F_L \sim (3,1,2)(-1/3),\quad E_R \sim (3,1,1)(-4/3),\quad N_R \sim
(3,1,1)(2/3),&\
\end{eqnarray}
where the standard leptons $\ell_L$, $e_R$ and $\nu_R$
are one of the colour components of $F_L$, $E_R$ and $N_R$, respectively.
This gauge structure can clearly support a
discrete symmetry between quarks and the generalized leptons. The most
straightforward possibility is the symmetry
\begin{equation}
Q_L \leftrightarrow F_L,\ \ u_R \leftrightarrow E_R,\ \ d_R
\leftrightarrow N_R,\ \  G^{\mu}_q \leftrightarrow G^{\mu}_{\ell},\ \
W^{\mu} \leftrightarrow W^{\mu},\ \ C^{\mu} \leftrightarrow -C^{\mu},
\label{ql-symmetry}
\end{equation}
where $G^{\mu}_q$, $G^{\mu}_{\ell}$, $W^{\mu}$ and $C^{\mu}$ are the
gauge bosons of SU(3)$_q$, SU(3)$_{\ell}$, SU(2)$_L$ and U(1)$_X$
respectively. Other varieties are also possible (see Ref.~\cite{FLV-9}
for a
complete discussion), but for definiteness we will concentrate on this
form of discrete q-$\ell$ symmetry in this paper.

The standard model is recovered as an effective
low-energy theory through the two-stage symmetry breaking chain
\begin{equation}
G_{q\ell} \to {\rm SU(2)}' \otimes G_{SM} \to {\rm SU(2)}' \otimes
{\rm SU(3)}_q \otimes {\rm U(1)}_Q,
\label{SSBchain}
\end{equation}
where SU(2)$'$ is an unbroken subgroup of leptonic colour
SU(3)$_{\ell}$. The discrete q-$\ell$ symmetry is broken at
the same time as leptonic colour in the simplest scenarios.
The weak hypercharge generator $Y$ of $G_{SM}$ is given by
\begin{equation}
Y = X + T/3,
\end{equation}
where $T \equiv {\rm diag}(-2,1,1)$ is one of the diagonal generators
of leptonic colour. Each standard lepton has a pair of exotic partners
through leptonic colour invariance. After the first stage of symmetry
breaking -- that is, $G_{q\ell} \to {\rm SU(2)}' \otimes G_{SM}$ -- the
standard leptons are identified as the $T=-2$ components of the leptonic
colour triplets, while the $T=1$ components form an SU(2)$'$ doublet of
exotic fermions with electric-charge $\pm 1/2$ called ``liptons.''
All of the particles which feel the residual SU(2)$'$ force can be made
heavy, apart from the gauge bosons of SU(2)$'$. Therefore, although the
unbroken gauge group at the electroweak scale is larger than $G_{SM}$,
the effective theory at this scale is still the SM.

\vskip 1 cm
\leftline{\bf 2.1 Higgs sector A}
\vskip 5 mm

The simplest way that the symmetry breaking chain of Eq.~(\ref{SSBchain})
can be induced is by choosing the Higgs sector,
\begin{equation}
\chi_1 \sim (1,3,1)(-2/3),\qquad \chi_2 \sim (3,1,1)(2/3),\qquad \phi
\sim (1,2,1)(1),
\label{higgsA}
\end{equation}
where $\chi_1 \leftrightarrow \chi_2$ and $\phi \leftrightarrow \phi^c
\equiv i\tau_2\phi^*$ under q-$\ell$ symmetry. Under the subgroup
SU(2)$' \otimes G_{SM}$ the transformation law for $\chi_2$ is
\begin{eqnarray}
\chi_2 & \to & (1,1,1)(0) \oplus (2,1,1)(1), \nonumber\\
\chi_2 & \to & \chi^0_2 \oplus \tilde{\chi}_2,
\label{chi}
\end{eqnarray}
where the second line establishes our nomenclature for the component
fields under the subgroup.
Therefore the first stage of symmetry breaking is induced by the vacuum
expectation value (VEV) pattern,
\begin{equation}
\langle\chi_1\rangle = 0,\quad \langle\chi^0_2\rangle = v \neq 0,\quad
\langle\tilde{\chi}_2\rangle = 0,\quad \langle\phi\rangle = 0,
\end{equation}
while the second stage (that is, standard electroweak breaking) is
induced by $\langle\phi\rangle = u \neq 0$.\footnote{Note that an SU(3)
transformation can be used to bring any VEV for $\chi_2$ into the
$\langle\chi^0_2\rangle = v \neq 0$,
$\langle\tilde{\chi}_2\rangle = 0$ form.}
The hierarchy $v \gg u$ is
required in order to satisfy phenomenological bounds.

All of the Higgs multiplets in Eq.~(\ref{higgsA}) couple to fermions
through Yukawa interactions. The Lagrangian is
\begin{eqnarray}
{\cal L}_{\rm Yuk} & = & h_1 [\overline{(Q_L)^c} Q_L \chi_1 +
\overline{(F_L)^c}
F_L \chi_2] + h_2 [\overline{(u_R)^c} d_R \chi_1 + \overline{(E_R)^c} N_R
\chi_2] \nonumber\\
& + & \lambda_1 [\overline{Q}_L d_R \phi + \overline{F}_L N_R
\phi^c] + \lambda_2 [\overline{Q}_L u_R \phi^c + \overline{F}_L E_R
\phi] + {\rm H.c.}
\label{LYuk}
\end{eqnarray}
After the first stage of symmetry breaking, liptons gain masses $h_1 v$
and $h_2 v$ from these Yukawa coupling terms. Since $v \gg u$ this means
that the liptons will in general be much more massive than standard
leptons and quarks, provided that the Yukawa coupling constants
$h_{1,2}$ are not extremely small. The standard fermions of course gain
masses after the second stage of symmetry breaking from the usual $\phi$
Yukawa interactions. Note, however, that the discrete q-$\ell$ symmetry
imposes the tree-level mass relations
\begin{equation}
m_u = m_e,\qquad m_d = m^{\rm Dirac}_{\nu},
\label{massrelations}
\end{equation}
which are not phenomenologically acceptable. Radiative corrections will
alter these mass relations, but not enough to make them tenable.

We therefore see that the symmetry breaking pattern we desire can be
induced by the simple Higgs sector of Eq.~(\ref{higgsA}). This scheme has
the desirable property that liptons are in general expected to be much
heavier than leptons and quarks. However, it has the undesirable mass
relations of Eq.~(\ref{massrelations}). The next Higgs sector we examine
retains the desirable features of this prototype but improves on the
unsuccessful mass relations.

\vskip 1 cm
\leftline{\bf 2.2 Higgs sector B}
\vskip 5 mm
\leftline{\it 2.2.1 Introduction}
\vskip 5 mm

The amended Higgs sector consists of $\chi_1$, $\chi_2$ plus two
electroweak Higgs doublets
\begin{equation}
\phi_1 \sim (1,1,2)(1) = \left( \begin{array}{c} \phi^+_1 \\ \phi^0_1
\end{array} \right)
\quad {\rm and}\quad \phi_2 \sim (1,1,2)(-1) = \left(
\begin{array}{c} \phi_2^0 \\ \phi_2^- \end{array} \right),
\end{equation}
which interchange under the q-$\ell$
discrete symmetry. The symmetry breaking pattern is the same as in
Eq.~(\ref{SSBchain}), with both electroweak doublets in general
participating in breaking the electroweak symmetry: $\langle\phi_1\rangle
= u_1$ and $\langle\phi_2\rangle = u_2$.

The Yukawa Lagrangian for the $\chi$ multiplets is obviously the same as
for Higgs sector A, but the $\phi$ interactions are different. There are
actually two different models which use Higgs sector B. These models are
distinguished by the way the charge conjugates of $\phi_1$ and $\phi_2$
behave under q-$\ell$ discrete symmetry. This in turn leads to two
different electroweak Yukawa Lagrangians and also to two different Higgs
potentials. The existence of two models using Higgs sector B is a subtle
point which was overlooked in previous papers \cite{FLV-2}.

\vskip 1 cm
\leftline{\it 2.2.2 Model 1}
\vskip 5 mm

Consider the charge conjugate Higgs fields $\phi^c_1$ and $\phi^c_2$
where
\begin{equation}
\phi^c_1 \equiv i\tau_2\phi_1^* = \left( \begin{array}{c} \phi_1^{0*}
\\ -\phi_1^- \end{array} \right)
\quad {\rm and}\quad \phi^c_2 \equiv
i\tau_2\phi^*_2 = \left( \begin{array}{c} \phi_2^+ \\ -\phi_2^{0*}
\end{array} \right).
\label{phic}
\end{equation}
In Model 1, the action of the q-$\ell$ discrete
symmetry on the Higgs doublets is
\begin{equation}
\phi_1 \leftrightarrow \phi_2,\quad {\rm and}\quad \phi_1^c
\leftrightarrow \phi_2^c,
\label{phic-interchange}
\end{equation}
which in terms of weak-isospin components has to be interpreted to
mean that components of like weak-isospin interchange:
\begin{equation}
\phi_1^+ \leftrightarrow \phi_2^0,\quad \phi_1^0 \leftrightarrow
\phi_2^-,\quad \phi_1^{0*} \leftrightarrow \phi_2^+,\quad
\phi_1^- \leftrightarrow \phi_2^{0*}.
\label{comp-interchange1}
\end{equation}
This last set of interchanges commutes with complex conjugation,
which is a necessary condition to maintain the invariance of the
kinetic energy terms for $\phi_{1,2}$. The alternative possibility
that unlike weak-isospin components interchange is not tenable,
because of the minus signs appearing in the definitions of the
charge conjugate doublets. For instance, $\phi_1^0 \leftrightarrow
\phi_2^0$ would have to be accompanied by $\phi_1^{0*} \leftrightarrow
-\phi_2^{0*}$ according to Eqs.~(\ref{phic}) and
(\ref{phic-interchange}). Since this does not preserve invariance of the
kinetic energy Lagrangian, this is not an allowed discrete symmetry and
so we must interpret Eq.~(\ref{phic-interchange}) as implying
Eq.~(\ref{comp-interchange1}) at the weak-isospin component level. We
similarly interpret the fermionic transformation $Q_L \leftrightarrow
F_L$ as implying
\begin{equation}
u_L \leftrightarrow N_L\quad {\rm and}\quad d_L \leftrightarrow E_L.
\label{ferm-interchange1}
\end{equation}
Model 1 therefore has the curious feature that the left- and
right-handed components of the quarks and leptons interchange in exactly
opposite ways, as can be seen by comparing
Eqs.~(\ref{ferm-interchange1}) and (\ref{ql-symmetry}). Another way to
put this is that the Model 1 discrete symmetry is chiral.

The Yukawa interactions for Model 1 are given by
\begin{eqnarray}
{\cal L}_{\rm Yuk} & = & \Lambda_1 [\overline{Q}_L d_R \phi_1 +
\overline{F}_L N_R \phi_2] + \Lambda'_1 [\overline{Q}_L d_R \phi_2^c +
\overline{F}_L N_R \phi_1^c] \nonumber\\
& + & \Lambda_2 [\overline{Q}_L u_R \phi_1^c + \overline{F}_L E_R
\phi_2^c] + \Lambda'_2 [\overline{Q}_L u_R \phi_2 + \overline{F}_L E_R
\phi_1] + {\rm H.c.}
\label{LYuk1}
\end{eqnarray}
The quark and lepton mass matrices are thus\footnote{After the first
stage of symmetry breaking, the lipton partners of the standard leptons
become heavy, and they will play no role in the rest of this paper. We
will alternate between the notation $e$ and $E$, and between
$\nu$ and $N$ whenever we find it convenient.}
\begin{eqnarray}
& m_u = \Lambda_2 u^*_1 + \Lambda'_2 u_2,\qquad m_e = -\Lambda_2 u^*_2 +
\Lambda'_2 u_1 &\ \nonumber\\
& m_d = \Lambda_1 u_1 - \Lambda_1' u^*_2,\qquad m^{\rm Dirac}_{\nu} =
\Lambda_1 u_2 + \Lambda'_1 u^*_1, &\
\label{qlmasses1}
\end{eqnarray}
and we see that the unsuccessful mass relations
that are unavoidable with Higgs sector A do not apply in general. It is
actually interesting to note that the relations $m_u = m_e$ and $m_d =
m_{\nu}$ hold for all choices of the $\Lambda$'s only if $u_1 = u_2 =
0$, that is, when the electroweak symmetry is unbroken. This is
consistent with Eq.(\ref{comp-interchange1}) because we see that q-$\ell$
discrete symmetry interchanges neutral Higgs bosons with charged Higgs
bosons. Therefore, the electroweak symmetry breaking VEVs $u_1$ and
$u_2$ for the neutral Higgs bosons
necessarily also break the discrete symmetry, and this is why the
tree-level fermionic mass relations are
necessarily violated in this model.\footnote{Of
course, the discrete symmetry is dominantly broken by the VEV for $\chi_2$.
The discrete symmetry breaking we are talking about in this section is
an additional contribution which comes from the $\phi$ bosons.}
Our previous observation that the left- and right-handed projections of
quarks and leptons transform in exactly opposite ways is of course also
consistent with the necessary violation of the fermionic mass
relations. [We will
see that Higgs sector A is similar to Model 2 to be presented below, in
that the electroweak symmetry breaking VEV('s) do {\it not}
inevitably also break q-$\ell$ symmetry.]

\vskip 1 cm
\leftline{\it 2.2.3 Model 2}
\vskip 5 mm

Model 2 is defined to obey the discrete symmetry
\begin{equation}
\phi_1 \leftrightarrow \phi_2,\quad {\rm and}\quad \phi_1^c
\leftrightarrow -\phi_2^c,
\end{equation}
which in component form has to mean that
\begin{equation}
\phi_1^+ \leftrightarrow \phi_2^-,\quad \phi_1^0 \leftrightarrow
\phi_2^0,\quad \phi_1^- \leftrightarrow \phi_2^+,\quad \phi_1^{0*}
\leftrightarrow \phi_2^{0*}.
\label{comp-interchange2}
\end{equation}
By contrast to Model 1, the components of unlike weak-isospin
interchange here. Note, of course, that the transformations in
Eq.~(\ref{comp-interchange2}) commute with complex conjugation and thus
leave the kinetic energy terms invariant. Similarly, the $Q_L
\leftrightarrow F_L$ interchange is interpreted to mean
\begin{equation}
u_L \leftrightarrow E_L\quad {\rm and}\quad d_L \leftrightarrow N_L,
\end{equation}
in Model 2. By contrast to Model 1 therefore, left- and right-handed
projections of the fermions transform in identical ways under the
discrete symmetry (in other words the discrete symmetry is
vector-like).\footnote{There is an alternative way to explain the
discrete symmetry used in Model 2. We present it here because it makes
no explicit reference to weak-isospin components, and thus should help
clarify what we mean by the above symmetry. The symmetry is equivalent
to $\phi_1 \to \tau_1\phi_2$, $F_L \to \tau_1 Q_L$,
$C^{\mu} \to -C^{\mu}$ and $\tau_i W^{\mu}_i \to \tau_1 \tau_i
W^{\mu}_i \tau_1$, where $\tau_1$ is the first Pauli matrix. Under these
transformations, $D^{\mu}\phi_1 \to \tau_1 D^{\mu}\phi_2$ where
$D^{\mu}$ is the gauge-covariant derivative. Therefore the
gauge--kinetic-energy terms for the two Higgs doublets interchange under
the action of the symmetry. One can easily check that the
gauge-invariant kinetic-energy terms for the fermions and for the gauge
bosons are also invariant.
This establishes that the discrete symmetry
of Model 2 is well-defined. The $\tau_1$ matrix effectively tells us
that components of unlike weak-isospin transform into each other here.
Note also that the $W$-boson transformation above equates to $W^{\mu}_3
\to -W^{\mu}_3$ and $W^{+\mu} \to W^{-\mu}$ at the component level.
We thank H. Lew for alerting us to this.}

The Yukawa Lagrangian for Model 2 is obtained from that of Model 1 by
the substitution $\phi_2^c \to -\phi_2^c$:
\begin{eqnarray}
{\cal L}_{\rm Yuk} & = & \Lambda_1 [\overline{Q}_L d_R \phi_1 +
\overline{F}_L N_R \phi_2] + \Lambda'_1 [-\overline{Q}_L d_R \phi_2^c +
\overline{F}_L N_R \phi_1^c] \nonumber\\
& + & \Lambda_2 [\overline{Q}_L u_R \phi_1^c - \overline{F}_L E_R
\phi_2^c] + \Lambda'_2 [\overline{Q}_L u_R \phi_2 + \overline{F}_L E_R
\phi_1] + {\rm H.c.}
\label{LYuk2}
\end{eqnarray}
The quark and lepton mass matrices are thus
\begin{eqnarray}
& m_u = \Lambda_2 u^*_1 + \Lambda'_2 u_2,\qquad m_e = \Lambda_2 u^*_2 +
\Lambda'_2 u_1 &\ \nonumber\\
& m_d = \Lambda_1 u_1 + \Lambda_1' u^*_2,\qquad m^{\rm Dirac}_{\nu} =
\Lambda_1 u_2 + \Lambda'_1 u^*_1. &\
\label{qlmasses2}
\end{eqnarray}
Once again, the bad mass relations of Eq.~(\ref{massrelations}) are in
general violated. Note, however, that the mass relations will be
reinstated in Model 2 if $u_1 = u_2 \neq 0$. This is consistent with the
transformation laws in Eq.~(\ref{comp-interchange2}) since the neutral
Higgs bosons interchange. If the two VEVs are equal, then discrete
q-$\ell$ symmetry is clearly not broken during electroweak symmetry
breaking, and so the tree-level fermionic mass relations ensue.

For the sake of completeness we note that Higgs sector A behaves in
a similar way to Model 2. The transformation
$\phi \leftrightarrow \phi^c$ has to be interpreted as implying $\phi^0
\leftrightarrow \phi^{0*}$ and $\phi^+ \leftrightarrow -\phi^-$. Since
the phase of the VEV $u$ is unobservable, q-$\ell$ symmetry is not
broken during electroweak symmetry breaking and so the bad mass
relations follow.

\vskip 1 cm
\leftline{\it 2.2.4 Discussion}
\vskip 5 mm

Although both Models 1 and 2 using Higgs sector B
have no mass relation problem, they also have no predictive power for
masses. This is perhaps unfortunate, since we would prefer to have an
enlarged symmetry like discrete q-$\ell$ symmetry actually increase the
predictive power for the masses of quarks and leptons. Higgs
sector A is predictive, but the predictions are wrong. Another response
to the problem posed by Higgs sector A would therefore be to look for
some other modification of it which maintained its predictive power
for masses but
this time with correct predictions. No such modification is known at
present, but its desirability motivates that an on-going search be
maintained.

Because the use of Higgs sector B is the simplest way to avoid the
fermion mass relation problem, it will be the principal focus of study
in this paper (see Sec.III).
The main new qualitative result to be presented is that
{\it discrete q-$\ell$ symmetry continues to furnish us with more predictive
power, even when all trace of it has disappeared from the fermion mass
spectrum.} We will see that quark-lepton partnership in both Models 1
and 2 is manifested in
the Yukawa interactions between fermions and physical Higgs bosons. In
particular, we will see that the flavour-changing neutral Higgs-boson
term for a given fermion is proportional to the mass matrix of its
partner under the discrete symmetry.

\vskip 1 cm
\leftline{\bf 2.3 Other Higgs sectors}
\vskip 5 mm

There are several other interesting Higgs sectors one can use. For
instance if one wishes to address the issue of small neutrino masses,
one may introduce a see-saw mechanism \cite{seesaw-8}
by the introduction of the Higgs multiplets \cite{FL-1},
\begin{equation}
\Delta_1 \sim (1,\overline{6},1)(4/3)\quad {\rm and}\quad \Delta_2 \sim
(\overline{6},1,1)(-4/3),
\end{equation}
where $\Delta_1$ and $\Delta_2$ interchange under q-$\ell$ symmetry.
A large VEV for the neutral component of $\Delta_2$ induces large
Majorana masses for the right-handed neutrinos, thus producing the
see-saw phenomenon. Either Higgs sector A or Higgs sector B can be
augmented by the introduction of these antisextets, and we will call the
resulting Higgs sector generically as ``Higgs sector C.''

Another Higgs field of some interest is a real gauge singlet $\sigma$
which is odd under q-$\ell$ symmetry ($\sigma \to -\sigma$). The purpose
of $\sigma$ would be to separate the scales of leptonic colour and
discrete symmetry breaking. A motivation for this might be cosmology,
because such a scenario allows one to break the discrete symmetry before
an inflationary epoch in the Hot Big Bang picture, while leptonic colour
could be broken after inflation. This can be used to render innocuous
the cosmological domain walls formed during the q-$\ell$ symmetry
breaking phase transition, while retaining exact leptonic colour down to
TeV-scale energies \cite{LV-5}. Any of Higgs sectors
A, B or C can extended by introducing $\sigma$.

Finally, we comment that
the gauge group $G_{q\ell}$ is but the simplest symmetry which supports
a discrete q-$\ell$ symmetry. An interesting extension is provided by
the gauge group $G_{q\ell LR}$ where
\begin{equation}
G_{q\ell LR} = {\rm SU(3)}_{\ell} \otimes {\rm SU(3)}_q \otimes
{\rm SU(2)}_L \otimes {\rm SU(2)}_R \otimes {\rm U(1)}_V.
\end{equation}
This model can support left-right discrete symmetry as well as q-$\ell$
symmetry \cite{FL-10}.
An even simpler extension sees only the U(1) subgroup of
SU(2)$_R$ gauged. Any gauge extension like this will also require an
extended Higgs sector. We will not probe this issue any further in this
paper, but have mentioned it here for the sake of completeness.

\vskip 1 cm
\leftline{\large \bf 3. Study of a realistic Higgs sector}
\vskip 5 mm

We now study the two models using Higgs sector B in more detail.
We will address the
following issues: (i) the construction and minimization of the Higgs
potentials, (ii) the spectrum and Yukawa couplings of the physical
Higgs bosons,
and (iii) the phenomenological bounds obtained from tree-level
flavour-changing neutral
Higgs boson effects in the light $\phi_{1,2}$ sector. Models 1 and 2
differ from each other in important ways, and so we will examine them
separately.

\vskip 1 cm
\leftline{\bf 3.1 Model 1}
\vskip 5 mm

The first issue is the minimization of the Higgs potential: we have to
check that the desired symmetry breaking pattern,
\begin{equation}
\langle\phi_1\rangle = \left( \begin{array}{c} 0 \\ u_1 \end{array}
\right),\quad \langle\phi_2\rangle = \left( \begin{array}{c} u_2 e^{i\xi}
\\ 0 \end{array} \right),\quad
\langle\chi_1\rangle = 0,
\quad \langle\chi_2\rangle =
\left( \begin{array}{c} v \\ 0 \\ 0 \end{array} \right),
\label{vev-pattern}
\end{equation}
is possible. In this equation we have chosen $u_1, u_2, v > 0$ by a
phase convention, and we have also
taken the possible phase angle $\xi$ to reside with $\phi_2$.

In order to construct the Higgs potential, we first write down all
gauge-invariant terms with definite transformation
properties under q-$\ell$ symmetry that are quadratic in the Higgs boson
fields. The terms $E$ that are even ($E \to E$) under the discrete
symmetry are,
\begin{equation}
E_1 \equiv \phi_1^{\dagger}\phi_1 + \phi_2^{\dagger}\phi_2\quad {\rm
and}\quad E_2 \equiv \chi_1^{\dagger}\chi_1 + \chi_2^{\dagger}\chi_2.
\end{equation}
The terms $O$ that are odd ($O \to -O$) are,
\begin{equation}
O_1 \equiv \phi_1^{\dagger}\phi_1 - \phi_2^{\dagger}\phi_2,\quad
O_2 \equiv \chi_1^{\dagger}\chi_1 - \chi_2^{\dagger}\chi_2,\quad
O_3 \equiv \phi_1\phi_2,\quad {\rm and}\quad O_3^{\dagger} \equiv
(\phi_1\phi_2)^{\dagger},
\end{equation}
where
\begin{equation}
\phi_1\phi_2 \equiv \phi^{\top}_1 i\tau_2\phi_2 = \phi_1^+\phi_2^- -
\phi_1^0\phi_2^0.
\end{equation}
The Higgs potential $V$ is obtained from these terms as follows:
\begin{eqnarray}
V & = & \mu^2_{\phi} E_1 + \mu^2_{\chi} E_2
 +  \lambda_1 E_1^2 + \lambda_2 E_2^2\nonumber\\
& + & \lambda_3 O_1^2 + \lambda_4 O_2^2 + [\lambda_5 O_3^2 +
{\rm H.c.}]\nonumber\\
& + & \lambda_6 E_1 E_2 + \lambda_7 O_1 O_2 + [\lambda_8 O_1 O_3 + {\rm
H.c.}]\nonumber\\
& + & [\lambda_9 O_2 O_3 + {\rm H.c.}] + \lambda_{10} O_3 O_3^{\dagger}
\end{eqnarray}
where $\mu^2_{\phi,\chi}$, $\lambda_{1-4,6,7,10}$ are real numbers and
$\lambda_{5,8,9}$ are in general complex. There are no quartic terms that
cannot be written down as the product of $E$s and $O$s.

The minimization equations for the Higgs potential written in this form
are not particularly enlightening, and it is difficult to tell if minima
are local or global. The terms $\mu^2_{\phi,\chi}$ and
$\lambda_{1-4,6}$ can actually be written
in a much more useful form. We will call the resulting partial Higgs potential
$V_{\rm large}$, and it is given by
\begin{eqnarray}
V_{\rm large} & = & \lambda'_1 (\phi_1^{\dagger}\phi_1 +
\phi_2^{\dagger}\phi_2 - u_1^2 - u_2^2)^2 + \lambda'_2
(\chi_1^{\dagger}\chi_1 + \chi_2^{\dagger}\chi_2 - v^2)^2\nonumber\\
& + & \lambda'_6 (\phi_1^{\dagger}\phi_1 +
\phi_2^{\dagger}\phi_2 + \chi_1^{\dagger}\chi_1 + \chi_2^{\dagger}\chi_2
- u_1^2 - u_2^2 - v^2)^2\nonumber\\
& + & \lambda'_4 (\chi_1^{\dagger}\chi_1)(\chi_2^{\dagger}\chi_2) +
\lambda'_3 [(\phi_1^{\dagger}\phi_1)(\phi_2^{\dagger}\phi_2) -
(\phi_1\phi_2)(\phi_1\phi_2)^{\dagger}].
\label{Vlarge}
\end{eqnarray}
The remaining terms in the full Higgs potential are assembled into a
contribution called $V_{\rm small}$ so that $V = V_{\rm large} +
V_{\rm small}$.

Consider the parameter space region given by $\lambda'_{1-4,6} > 0$.
The first three terms in $V_{\rm large}$ are clearly minimized by taking
$\langle\phi_1^{\dagger}\phi_1\rangle +
\langle\phi_2^{\dagger}\phi_2\rangle = u_1^2 + u_2^2$ and
$\langle\chi_1^{\dagger}\chi_1\rangle
+ \langle\chi_2^{\dagger}\chi_2\rangle = v^2$. The fourth term
is minimized if either $\langle\chi_1\rangle = 0$ or $\langle\chi_2\rangle
= 0$. Without loss of generality we may take $\langle\chi_1\rangle = 0$,
thereby being consistent with our previous exposition. The fourth term
thus guarantees leptonic colour breaking, q-$\ell$ symmetry breaking and
the preservation of quark colour as an exact symmetry. The last term
guarantees that electromagnetic gauge invariance remains exact. The
argument goes like this: We start by using a weak-isospin rotation to
transform to the basis where $\langle\phi_1^{\pm}\rangle = 0$.
This however does not
ensure that $\langle\phi_2^{\pm}\rangle = 0$.
But the last term in Eq.~(\ref{Vlarge})
at the minimum is then just $\lambda'_3
\langle\phi_2^-\phi_2^+\rangle|\langle\phi_1^0\rangle|^2$. Since
$\lambda'_3 > 0$, then either $\langle\phi_2^{\pm}\rangle = 0$ or
$\langle\phi_1^0\rangle = 0$, and so we choose the former.

We have thus shown that the VEV pattern of Eq.~(\ref{vev-pattern})
arises when all the $\lambda'$ coupling constants in $V_{\rm large}$ are
positive, provided that the omitted terms in $V_{\rm small}$ are small
enough. We have not shown that this is the only region of parameter
space that will do, only that it is an example of a suitable region.
This is perhaps fortunate, because this region has a serious drawback:
the existence of a light pseudo-Goldstone boson. The point is that
$V_{\rm large}$ is invariant under independent phase rotations for
$\chi_2$, $\phi_1$ and $\phi_2$ that are all spontaneously broken. Two
of the resulting Goldstone bosons are eaten, but one remains as a light
physical boson. It will pick up some mass from $V_{\rm small}$
and via radiative corrections from the Yukawa Lagrangian, but the
fear is that it will be light enough to mediate unacceptably large
neutral flavour-changing processes.

This problem has arisen because the $\phi_1\phi_2$ combination is odd
under the discrete symmetry, and thus cannot appear in Higgs potential terms
of the form (positive number)$\times (\phi_1\phi_2 + u_1 u_2
\cos\xi)^2$. Terms like $(\phi_1\phi_2)^2$ eliminate the spurious phase
invariance, but they seemingly cannot be written in a manifestly useful
way for purposes of easy minimization while at the same time respecting
the discrete symmetry. However, we can easily convince ourselves that a
pseudo-Goldstone boson does not necessarily always accompany our
required VEV pattern. Let us write down an {\it effective} Higgs
potential for $\phi_1$ and $\phi_2$ after a nonzero VEV for $\chi_2$ has
already spontaneously broken the discrete symmetry. We do this by
allowing the soft discrete symmetry breaking quadratic terms
$\phi_1\phi_2$ and $\phi_1^{\dagger}\phi_1 - \phi_2^{\dagger}\phi_2$
to appear in
our effective potential. The most general form is
\begin{eqnarray}
V_{\rm eff} & = & \lambda''_1 (\phi_1^{\dagger}\phi_1 +
\phi_2^{\dagger}\phi_2 - u_1^2 - u_2^2)^2\nonumber\\
& + & \lambda''_3 [(\phi_1^{\dagger}\phi_1)(\phi_2^{\dagger}\phi_2) -
(\phi_1\phi_2)(\phi_1\phi_2)^{\dagger}]
 +  \lambda''_7 (\phi_1^{\dagger}\phi_1 -
\phi_2^{\dagger}\phi_2 - u_1^2 + u_2^2)^2\nonumber\\
& + & \lambda''_5 [(\phi_1\phi_2) + (\phi_1\phi_2)^{\dagger} +
2u_1u_2\cos\xi]^2\nonumber\\
& + & \tilde{\lambda}''_5 [i(\phi_1\phi_2) -
i(\phi_1\phi_2)^{\dagger} - 2u_1u_2\sin\xi]^2\nonumber\\
& + & \lambda''_8[\phi_1^{\dagger}\phi_1 - \phi_2^{\dagger}\phi_2 +
(\phi_1\phi_2) + (\phi_1\phi_2)^{\dagger} - u_1^2 + u_2^2
+ 2u_1u_2\cos\xi]^2\nonumber\\
& + & \tilde{\lambda}''_8[\phi_1^{\dagger}\phi_1 - \phi_2^{\dagger}\phi_2 +
i(\phi_1\phi_2) - i(\phi_1\phi_2)^{\dagger} - u_1^2 + u_2^2
- 2u_1u_2\sin\xi]^2\nonumber\\
& + & \lambda''_{10} [(\phi_1\phi_2) + (\phi_1\phi_2)^{\dagger}\nonumber\\
&\ &\qquad\qquad +\ i(\phi_1\phi_2) -
i(\phi_1\phi_2)^{\dagger} +  2u_1u_2\cos\xi -
2u_1u_2\sin\xi]^2.
\end{eqnarray}
If all the $\lambda''$ parameters are positive, then the required
pattern of symmetry breaking follows. Furthermore, there is no reason to
make the terms breaking the spurious phase invariance small, and so the
putative pseudo-Goldstone boson is eliminated. This result shows that
some of the terms in $V_{\rm small}$, which are putatively small,
can actually be large enough to solve this
pseudo-problem without inducing the
spontaneous breaking of electromagnetic gauge invariance, or otherwise
spoiling our desired symmetry breaking pattern.

We now exhibit the physical Higgs fields. Writing
\begin{equation}
\phi_1^0 = u_1 + {{h_1+i\eta_1} \over \sqrt{2}},\quad \phi_2^0 = u_2 +
{{h_2+i\eta_2} \over \sqrt{2}},\quad \chi_2^0 = v + {{H+iE} \over
\sqrt{2}},
\end{equation}
where for simplicity we have put the CP-violating phase $\xi$ to zero,
we identify the unphysical neutral Goldstone boson fields as
\begin{equation}
g^0 = {{u_2\eta_2 - u_1\eta_1} \over \sqrt{u_1^2 + u_2^2}}\quad
{\rm and}\quad E,
\end{equation}
where $g^0$ is the field eaten by the $Z^0$.
The field $E$ is eaten by a $Z'$ boson arising from the spontaneous
breakdown of leptonic colour. The field we called $\tilde{\chi}_2$ in
Eq.~(\ref{chi}) is also eaten when leptonic colour breaks.
The unphysical charged Goldstone bosons are
\begin{equation}
g^{\pm} \equiv {{u_2\phi_2^{\pm} - u_1\phi_1^{\pm}} \over \sqrt{u_1^2 +
u_2^2}},
\end{equation}
and they are of course eaten by $W^{\pm}$.
{}From now on we will work in unitary gauge, so that these unphysical
fields will simply be set to zero in the Yukawa Lagrangians.

The physical Higgs fields consist of the charged field $H^{\pm}$
orthogonal to $g^{\pm}$, where
\begin{equation}
H^{\pm} \equiv {{u_2\phi_1^{\pm} + u_1\phi_2^{\pm}} \over \sqrt{u_1^2 +
u_2^2}};
\end{equation}
the CP-odd field $\eta$ orthogonal to $g^0$, where
\begin{equation}
\eta \equiv {{u_2\eta_1 + u_1\eta_2} \over \sqrt{u_1^2 + u_2^2}};
\end{equation}
and three CP-even fields whose mass eigenstates are linear combinations
of $h_1$, $h_2$ and $H$. Now, in most of parameter
space the field $H$ mixes very little with the fields $h_{1,2}$,
because the scale of leptonic colour breaking has to be significantly
higher than the electroweak scale. We will concentrate on this large
region of parameter space in the rest of this paper. We therefore
approximately write the mass eigenstate fields as $h'_{1,2}$ and $H$,
where
\begin{equation}
\left( \begin{array}{c} h'_1 \\ h'_2 \end{array} \right) =
\left( \begin{array}{cc} \cos\phi & \sin\phi \\ -\sin\phi & \cos\phi
\end{array} \right) \left( \begin{array}{c} h_1 \\ h_2 \end{array}
\right),
\end{equation}
for some mixing angle $\phi$. We can of course relate $\phi$ to the
parameters in the Higgs potential, but we will not need to know this
expression.

The most useful way to write the Yukawa Lagrangian is to replace the
$\Lambda$ parameters in Eq.~(\ref{LYuk1}) by the mass matrices through
Eq.~(\ref{qlmasses1}). We then obtain that
\begin{eqnarray}
{\cal L}_{\rm Yuk} & = & {1 \over u} \bar{Q}_L (m_u \Phi_1 + m_e \Phi_2)
u_R + {1 \over u} \bar{F}_L (m_u \Phi_2^c - m_e\Phi_1^c) E_R\nonumber\\
& + & {1 \over u} \bar{Q}_L (m^{\rm Dirac}_{\nu} \Phi_2^c
- m_d \Phi_1^c) d_R +
{1 \over u} \bar{F}_L (m^{\rm Dirac}_{\nu} \Phi_1
+ m_d \Phi_2) N_R + {\rm H.c.}
\label{LYuk1-primed}
\end{eqnarray}
where $u \equiv \sqrt{u_1^2 + u_2^2}$ and
\begin{eqnarray}
\Phi_1 & \equiv & {{u_2\phi_2 + u_1\phi_1^c} \over u} =
\left( \begin{array}{c}
u + {{u_1 h_1 + u_2 h_2} \over {\sqrt{2}u}} + i {g^0 \over \sqrt{2}} \\
g^- \end{array} \right),\\
\Phi_2 & \equiv & {{u_1\phi_2 - u_2\phi_1^c} \over u} =
\left( \begin{array}{c}
{{u_1 h_2 - u_2 h_1} \over {\sqrt{2}u}} + i {\eta \over \sqrt{2}} \\
H^- \end{array} \right),
\label{Phi2}
\end{eqnarray}
and $\Phi^c_{1,2} \equiv i\tau_2\Phi^*_{1,2}$. Under q-$\ell$
symmetry,
\begin{equation}
\Phi_1 \leftrightarrow \Phi_2^c\quad {\rm and}\quad
\Phi_2 \leftrightarrow -\Phi_1^c,
\end{equation}
and the electroweak symmetry breaking VEVs are
$\langle\Phi_1\rangle = u$ and $\langle\Phi_2\rangle = 0$.

By using Eq.~(\ref{Phi2}),
the Yukawa Lagrangians involving the physical mass eigenstate fields
$H^{\pm}$ and $\eta$ can be easily read off Eq.~(\ref{LYuk1-primed}).
They are,
\begin{eqnarray}
{\cal L}_{\rm Yuk}^+ & = & {1 \over u} \bar{d}_L m_E u_R H^- + {1 \over
u} \bar{u}_L m^{\rm Dirac}_{\nu} d_R H^+\nonumber\\
& + & {1 \over u} \bar{N}_L m_u E_R H^+ + {1 \over u} \bar{E}_L m_d
N_R H^- + {\rm H.c.}
\label{LYuk1-charged}
\end{eqnarray}
and
\begin{eqnarray}
{\cal L}_{\rm Yuk}^{\eta} & = & {i \over {\sqrt{2}u}} \bar{u}_L m_E u_R
\eta + {i \over {\sqrt{2}u}} \bar{E}_L m_u E_R \eta\nonumber\\
& + & {i \over {\sqrt{2}u}} \bar{d}_L m^{\rm Dirac}_{\nu} d_R \eta
+ {i \over {\sqrt{2}u}} \bar{N}_L m_d N_R \eta + {\rm H.c.}
\label{LYuk1-eta}
\end{eqnarray}
These Yukawa Lagrangians are extremely interesting, because the discrete
q-$\ell$ symmetry is seen to act in a highly non-trivial way:
{\it The Yukawa coupling constants for quarks (leptons) are proportional
to the mass matrices of the corresponding discrete symmetry
partner leptons (quarks).} This is a rather different situation from
the usual expectation that the Yukawa coupling constants for fermion $f$
should be proportional to the mass $m_f$ of that same fermion. Note in
particular that the down-quark flavour-changing neutral couplings of the
$\eta$ Higgs boson are proportional to the Dirac masses of the
neutrinos. Since neutrino Dirac masses are constrained to be very small,
we see that neutral flavour changing processes mediated by $\eta$ are
highly suppressed for down quarks. This means, most importantly, that no
useful bound is obtained from $K-\bar{K}$ mixing on the tree-level
flavour-changing $\eta$ couplings.\footnote{If we extend the Higgs
sector to accomodate the see-saw mechanism (see Higgs sector C above)
then of course this qualitative conclusion no longer holds, because
neutrino Dirac masses can then be large. We also point out that in the
Higgs sector B scenario we have no explanation for why the neutrino
Dirac masses should be so small. Our result is simply that {\it given}
tiny neutrino Dirac masses, {\it then} tiny down-quark sector
flavour-changing neutral
couplings follow.} The reader should also note that the fermion fields
in Eqs.~(\ref{LYuk1-charged}) and (\ref{LYuk1-eta}) are weak-interaction
eigenstates, not mass eigenstates. In the mass eigenstate basis, two
unitary diagonalization matrices would also appear in the Lagrangians, a
point we will return to later on.

Using the definition
\begin{equation}
\tan\omega \equiv u_2/u_1,
\end{equation}
we can write the interaction Lagrangian of the fermions with the mass
eigenstate CP-even bosons $h'_{1,2}$ as
\begin{eqnarray}
{\cal L}_{\rm Yuk}^{h} & = & {1 \over {\sqrt{2}u}} \bar{u}_L \Big[ m_u
\{\cos(\omega-\phi)h'_1 + \sin(\omega-\phi)h'_2\}\nonumber\\
&\ &\qquad\quad + m_e\{-\sin(\omega-\phi)h'_1 + \cos(\omega-\phi)h'_2\}
\Big] u_R \nonumber\\
& + & {1 \over {\sqrt{2}u}} \bar{E}_L \Big[ m_u
\{\sin(\omega-\phi)h'_1 - \cos(\omega-\phi)h'_2\}\nonumber\\
&\ &\qquad\quad + m_e\{\cos(\omega-\phi)h'_1 + \sin(\omega-\phi)h'_2\}
\Big] E_R\nonumber\\
& + & {1 \over {\sqrt{2}u}} \bar{d}_L \Big[ m_d
\{\cos(\omega-\phi)h'_1 + \sin(\omega-\phi)h'_2\}\nonumber\\
&\ &\qquad\quad + m^{\rm Dirac}_{\nu} \{\sin(\omega-\phi)h'_1
- \cos(\omega-\phi)h'_2\}
\Big] d_R\nonumber\\
& + & {1 \over {\sqrt{2}u}} \bar{N}_L \Big[ m_d
\{-\sin(\omega-\phi)h'_1 + \cos(\omega-\phi)h'_2\}\nonumber\\
&\ &\qquad\quad + m^{\rm Dirac}_{\nu}
\{\cos(\omega-\phi)h'_1 + \sin(\omega-\phi)h'_2\}
\Big] N_R + {\rm H.c.}
\end{eqnarray}
where again interaction eigenstates have been used for the fermions.
This Lagrangian is also very interesting, because in the fermion mass
eigenstate basis it is clear that the flavour-changing contributions
for a given fermion are always proportional to the mass matrix of its
q-$\ell$ partner (multiplied by diagonalization matrices). This result
is similar to that obtained for the boson $\eta$. Once again, the most
important consequence of this is the large suppression of down-quark
sector flavour-changing neutral Higgs effects because of their
proportionality to tiny neutrino Dirac mass matrices. Note, however,
that if we set $m^{\rm Dirac}_{\nu} = 0$
exactly, then {\it all} interactions between
$\eta$ and the down-quarks disappear, whereas the $h_{1,2}$ fields still
have interactions although they are now strictly flavour-diagonal.

Having made the important discovery that discrete q-$\ell$ symmetry
plays an interesting and important role in constraining the Yukawa
interactions of physical Higgs bosons and fermions in Model 1, we now
turn to a similar analysis of Model 2. We will then return to Model 1
when we come to examine phenomenological bounds on tree-level
flavour-changing neutral processes in a subsequent subsection.

\vskip 1 cm
\leftline{\bf 3.2 Model 2}
\vskip 5 mm

Happily, the analysis of the Higgs potential for Model 2 is much simpler
than for Model 1. This is because the combination $\phi_1\phi_2 \equiv
\phi_1^{\top}i\tau_2\phi_2$ is now even under the discrete q-$\ell$
symmetry. The even and odd quadratic combinations are now given by
\begin{equation}
E_1 \equiv \phi_1^{\dagger}\phi_1 + \phi_2^{\dagger}\phi_2,\quad
E_2 \equiv \chi_1^{\dagger}\chi_1 + \chi_2^{\dagger}\chi_2,\quad
E_3 \equiv \phi_1\phi_2,\quad E^{\dagger}_3 \equiv
(\phi_1\phi_2)^{\dagger}
\end{equation}
and
\begin{equation}
O_1 \equiv \phi_1^{\dagger}\phi_1 - \phi_2^{\dagger}\phi_2,\quad
O_2 \equiv \chi_1^{\dagger}\chi_1 - \chi_2^{\dagger}\chi_2
\end{equation}
respectively. As for Model 1, the Higgs potential $V$ can be written as
sums of $E$, $E^2$ and $O^2$ forms. However, this time we can write all
but one of these terms immediately in a form useful for answering
minimization questions. The result is
\begin{eqnarray}
V_{\rm large} & = & \lambda'_1 (\phi_1^{\dagger}\phi_1 +
\phi_2^{\dagger}\phi_2 - u_1^2 - u_2^2)^2 + \lambda'_2
(\chi_1^{\dagger}\chi_1 + \chi_2^{\dagger}\chi_2 - v^2)^2\nonumber\\
& + & \lambda'_3 [(\phi_1^{\dagger}\phi_1)(\phi_2^{\dagger}\phi_2) -
(\phi_1\phi_2)(\phi_1\phi_2)^{\dagger}]
+ \lambda'_4 (\chi_1^{\dagger}\chi_1)(\chi_2^{\dagger}\chi_2)\nonumber\\
& + & \lambda'_5 [\phi_1\phi_2 + (\phi_1\phi_2)^{\dagger} +
2u_1 u_2\cos\xi]^2\nonumber\\
& + & \tilde{\lambda}'_5 [i\phi_1\phi_2 - i(\phi_1\phi_2)^{\dagger} -
2u_1 u_2\sin\xi]^2\nonumber\\
& + & \lambda'_6 (\phi_1^{\dagger}\phi_1 +
\phi_2^{\dagger}\phi_2 + \chi_1^{\dagger}\chi_1 + \chi_2^{\dagger}\chi_2
- u_1^2 - u_2^2 - v^2)^2\nonumber\\
& + & \lambda'_8 [\phi_1^{\dagger}\phi_1 +
\phi_2^{\dagger}\phi_2 + \phi_1\phi_2 + (\phi_1\phi_2)^{\dagger} - u_1^2
- u_2^2 + 2u_1 u_2\cos\xi]^2\nonumber\\
& + & \tilde{\lambda}'_8 [\phi_1^{\dagger}\phi_1 +
\phi_2^{\dagger}\phi_2 + i\phi_1\phi_2 - i(\phi_1\phi_2)^{\dagger} - u_1^2
- u_2^2 - 2u_1 u_2\sin\xi]^2\nonumber\\
& + & \lambda'_9 [\chi_1^{\dagger}\chi_1 +
\chi_2^{\dagger}\chi_2 + \phi_1\phi_2 + (\phi_1\phi_2)^{\dagger}
+ 2u_1 u_2\cos\xi - v^2]^2\nonumber\\
& + & \tilde{\lambda}'_9 [\chi_1^{\dagger}\chi_1 +
\chi_2^{\dagger}\chi_2 + i\phi_1\phi_2 - i(\phi_1\phi_2)^{\dagger} -
2u_1 u_2\sin\xi - v^2]^2\nonumber\\
& + & \lambda'_{10} [\phi_1\phi_2 + (\phi_1\phi_2)^{\dagger}\nonumber\\
&\ &\qquad\qquad +
i\phi_1\phi_2 - i(\phi_1\phi_2)^{\dagger} + 2u_1 u_2\cos\xi - 2u_1
u_2\sin\xi]^2.
\end{eqnarray}
If we take all the $\lambda'$ parameters above to be positive, then the
correct symmetry breaking pattern is assured, provided that the one
omitted term, namely $V_7$ where
\begin{equation}
V_7 = \lambda_7 (\phi_1^{\dagger}\phi_1 -
\phi_2^{\dagger}\phi_2)(\chi_1^{\dagger}\chi_1 - \chi_2^{\dagger}\chi_2)
\end{equation}
is small enough, or innocuous enough. If we want to we can partially
incorporate this term in our analysis by writing an effective
soft-breaking term of the form $\lambda''_7[O_1 - u_1^2 + u_2^2]^2$
in an effective
potential after leptonic colour breakdown, but there is no practical
need to examine this term more closely.

The identification of physical and unphysical fields is exactly the same
as for Model 1. The Yukawa Lagrangians are, however, a little
different. The charged Higgs boson Yukawa Lagrangian is
\begin{eqnarray}
{\cal L}^+_{\rm Yuk} & = & {1 \over {u(u_2^2-u_1^2)}} \bar{d}_L
(2u_1 u_2 m_u - u^2 m_e) u_R H^-\nonumber\\
& + & {1 \over {u(u_2^2-u_1^2)}} \bar{N}_L
(u^2m_u - 2u_1 u_2 m_e) E_R H^+\nonumber\\
& + & {1 \over {u(u_2^2-u_1^2)}} \bar{u}_L
(- 2u_1 u_2 m_d + u^2 m^{\rm Dirac}_{\nu}) d_R H^+\nonumber\\
& + & {1 \over {u(u_2^2-u_1^2)}} \bar{E}_L
(- u^2m_d + 2u_1 u_2 m^{\rm Dirac}_{\nu}) N_R H^- + {\rm H.c.}
\end{eqnarray}
Note that the q-$\ell$ partnership is manifested in a more complicated
way for this Lagrangian compared with its analogue in Model 1.

The CP-odd neutral particle $\eta$ enjoys the following interactions:
\begin{eqnarray}
{\cal L}^{\eta}_{\rm Yuk} & = & {i \over {\sqrt{2}u(u_2^2-u_1^2)}}
\bar{u}_L (2u_1 u_2 m_u - u^2 m_e) u_R \eta\nonumber\\
& + & {i \over {\sqrt{2}u(u_2^2-u_1^2)}}
\bar{E}_L (u^2m_u - 2u_1 u_2 m_e) E_R \eta\nonumber\\
& + & {i \over {\sqrt{2}u(u_2^2-u_1^2)}} \bar{d}_L
(- 2u_1 u_2 m_d + u^2 m^{\rm Dirac}_{\nu}) d_R \eta\nonumber\\
& + & {i \over {\sqrt{2}u(u_2^2-u_1^2)}}
\bar{N}_L (- u^2m_u + 2u_1 u_2 m_e) N_R \eta + {\rm H.c.}
\label{LYuk2-eta}
\end{eqnarray}
As for Model 1, the flavour-changing interaction of $\eta$
for a given fermion class is proportional to the mass matrix of the
discrete symmetry partner of that fermion. Most importantly, the
down-quark neutral flavour-violating piece is once again
proportional to the Dirac mass matrix of the neutrino and is therefore very
small. A difference from Model 1 is that there is non-zero piece for a
given fermion proportional to the mass matrix of that same fermion (and
is thus diagonal in the mass eigenstate basis).

The CP-even mass eigenstate Higgs bosons $h'_{1,2}$ have an interaction
Lagrangian given by
\begin{eqnarray}
{\cal L}_{\rm Yuk}^{h} & = & {u \over {\sqrt{2}(u_2^2 - u_1^2)}}
\bar{u}_L \Big[ m_u
\{-\cos(\omega+\phi)h'_1 + \sin(\omega+\phi)h'_2\}\nonumber\\
&\ &\quad\quad + m_e\{\sin(\omega-\phi)h'_1 - \cos(\omega-\phi)h'_2\}
\Big] u_R \nonumber\\
& + & {u \over {\sqrt{2}(u_2^2 - u_1^2)}} \bar{E}_L \Big[ m_u
\{\sin(\omega-\phi)h'_1 - \cos(\omega-\phi)h'_2\}\nonumber\\
&\ &\quad\quad + m_e\{-\cos(\omega+\phi)h'_1 + \sin(\omega+\phi)h'_2\}
\Big] E_R\nonumber\\
& + & {u \over {\sqrt{2}(u_2^2-u_1^2)}} \bar{d}_L \Big[ m_d
\{-\cos(\omega+\phi)h'_1 + \sin(\omega+\phi)h'_2\}\nonumber\\
&\ &\quad\quad + m^{\rm Dirac}_{\nu}\{\sin(\omega-\phi)h'_1
- \cos(\omega-\phi)h'_2\}
\Big] d_R\nonumber\\
& + & {u \over {\sqrt{2}(u_2^2-u_1^2)}} \bar{N}_L \Big[ m_d
\{\sin(\omega-\phi)h'_1 - \cos(\omega-\phi)h'_2\}\nonumber\\
&\ &\quad\quad + m^{\rm Dirac}_{\nu}\{-\cos(\omega+\phi)h'_1
+ \sin(\omega+\phi)h'_2\}
\Big] N_R + {\rm H.c.}
\label{LYuk2-h}
\end{eqnarray}
As the reader can easily see, the flavour-changing interaction for
a given fermion class is proportional to the mass matrix
of its q-$\ell$ symmetry partner. Once again, down-quark sector neutral
flavour-violating processes are zero if the neutrino Dirac masses are
zero.

\vskip 1 cm
\leftline{\bf 3.3 Phenomenology}
\vskip 5 mm

In this section we will present an overview of the phenomenological
implications of Models 1 and 2. The main interest is on the
tree-level neutral flavour-changing effects mediated by $\eta$, $h'_1$
and $h'_2$. (We will not cover loop effects quantitatively in this paper;
we will be content to qualitatively discuss the most interesting of
these here. We hope to return to a more detailed phenomenological
analysis in future work.)

\vskip 1 cm
\leftline{\it 3.3.1 Tree-level neutral flavour-changing effects.}
\vskip 5 mm

Let the mass eigenstate fermion field $f$ be denoted by $f'$. We
introduce the left- and right-sector unitary diagonalization matrices
$V^f_{L,R}$ through,
\begin{equation}
f'_L \equiv V^f_L f_L,\quad {\rm and}\quad f'_R \equiv V^f_R f_R,
\end{equation}
where $f = u,d,e,\nu$. The corresponding diagonal mass matrices are given
by
\begin{equation}
m_f^{\rm diag} = V^f_L m_f V_R^{f\dagger}.
\end{equation}
We now rewrite the flavour-changing pieces of the neutral Higgs Yukawa
Lagrangians in terms of mass eigenstate fermions.

For Model 1 we obtain,
\begin{eqnarray}
{\cal L}^{\eta}_{\rm FC} & = &
{1 \over {\sqrt{2}u}} \bar{u}'_L V^u_L V_L^{e\dagger} m^{\rm diag}_e
V_R^e V_R^{u\dagger} u'_R \eta
+ {1 \over {\sqrt{2}u}} \bar{e}'_L V^e_L V_L^{u\dagger} m^{\rm diag}_u
V_R^u V_R^{e\dagger} e'_R \eta\nonumber\\
& + & {1 \over {\sqrt{2}u}} \bar{d}'_L V^d_L V_L^{\nu\dagger} m^{\rm
diag}_{\nu} V_R^{\nu} V_R^{d\dagger} d'_R \eta
+ {1 \over {\sqrt{2}u}} \bar{\nu}'_L V^{\nu}_L V_L^{d\dagger}
m^{\rm diag}_d V_R^d V_R^{\nu\dagger} {\nu}'_R \eta;
\end{eqnarray}
together with
\begin{eqnarray}
{\cal L}^h_{\rm FC} & = &
{1 \over {\sqrt{2}u}} \bar{u}'_L V^u_L V_L^{e\dagger} m^{\rm diag}_e
V_R^e V_R^{u\dagger} u'_R [\sin(\omega-\phi)h'_1 +
\cos(\omega-\phi)h'_2]\nonumber\\
& + & {1 \over {\sqrt{2}u}} \bar{e}'_L V^e_L V_L^{u\dagger} m^{\rm
diag}_u V_R^u V_R^{e\dagger} e'_R [\sin(\omega-\phi)h'_1 -
\cos(\omega-\phi)h'_2]\nonumber\\
& + & {1 \over {\sqrt{2}u}} \bar{d}'_L V^d_L V_L^{\nu\dagger} m^{\rm
diag}_{\nu} V_R^{\nu} V_R^{d\dagger} d'_R [\sin(\omega-\phi)h'_1 -
\cos(\omega-\phi)h'_2]\nonumber\\
& + & {1 \over {\sqrt{2}u}} \bar{\nu}'_L V^{\nu}_L V_L^{d\dagger} m^{\rm
diag}_d V_R^d V_R^{\nu\dagger} {\nu}'_R [-\sin(\omega-\phi)h'_1 +
\cos(\omega-\phi)h'_2],
\end{eqnarray}
for flavour-changing $\eta$ and $h'_{1,2}$ interactions respectively.
(Note that the two Lagrangians above also contain flavour-diagonal
terms.)
The corresponding Lagrangians for Model 2 are easily discerned
from Eqs.~(\ref{LYuk2-eta}) and (\ref{LYuk2-h}).
They can be obtained from the two Model 1 Lagrangians above
by making the substitution $u \to (u_2^2-u_1^2)/u$ and by changing some
of the plus and minus signs.

Let us make some qualitative observations: (i) As we have emphasized, if
the neutrino Dirac masses are zero, then there are no tree-level
down-quark sector interactions which change flavour. In this case,
there are also no neutrino-sector flavour-changing vertices, because we
are free to redefine the neutrino fields by use of the down-sector
diagonalization matrices. (ii) All the action is therefore in the
up-quark and charged-lepton sectors. Since the up-quark sector masses
are larger than corresponding charged-lepton masses, the largest
flavour-changing couplings will occur for charged leptons. In
particular, those couplings proportional to the large top-quark mass
will dominate, unless they happen to be suppressed by small mixing angles.
(iii) Our experience with the Kobayashi-Maskawa (KM)
matrix suggests that the
inter-generational mixing pattern for these flavour-changing
interactions should be hierarchical. It would therefore follow that the
large top-quark mass will have most influence on $\tau \to \mu$
conversions. Of course, a hierarchical mixing pattern is not inevitable,
but at this juncture it nevertheless represents the best guess, in our
opinion.

To get a feeling for the likely strength of these flavour-changing
transitions, let us assume that all of the mixing matrices $V$ follow the
qualitative form of the KM matrix, namely
\begin{equation}
V \sim \left( \begin{array}{ccc}
1 & \epsilon & \epsilon^3 \\
\epsilon & 1 & \epsilon^2 \\
\epsilon^3 & \epsilon^2 & 1
\end{array} \right)
\end{equation}
where $\epsilon$ is a small parameter, which for the KM matrix equals
about $0.2$. Note that this qualitative pattern is preserved when two
such matrices are multiplied together. For purposes of illustration, if
we take $\epsilon \simeq 0.1$, then for both up-quarks and charged
leptons we find that
\begin{equation}
V^2 m^{\rm diag} V^2 \sim
\left( \begin{array}{ccc}
m_1 & \epsilon m_2 & \epsilon^3 m_3 \\
\epsilon m_2 & m_2 & \epsilon^2 m_3 \\
\epsilon^3 m_3 & \epsilon^2 m_3 & m_3
\end{array} \right)
\end{equation}
where $m_{1,2,3}$ refers to the first, second and third generation mass
respectively. So, looking at charged-lepton transitions, we see that
$\tau \to \mu$ is proportional to $\epsilon^2 m_t \simeq 1$ GeV.
Observe that $\tau \to e$ is a further power of $\epsilon$
smaller, while $\mu \to e$
is driven by $\epsilon m_c$ which happens to be the same order of
magnitude as $\tau \to e$.

Let us look now at the specific process $\mu^- \to e^- e^+ e^-$, which
will in general be mediated by all the neutral Higgs bosons $\eta$ and
$h'_{1,2}$. We will assume that all of the these Higgs particles have
roughly the same mass $m_{\phi}$, and we will assume that no accidental
cancellations occur between the three tree-level Feynman graphs
contributing to this process. The partial decay width $\Gamma$
is then roughly given by
\begin{equation}
\Gamma \simeq 10^{-4} \times
\left({{\epsilon m_c m_u} \over m^2_{\phi}}\right)^2
{m^5_{\mu} \over u^4}.
\end{equation}
[This decay rate is calculated within Model 1. The Model 2 estimate is
of exactly the same form, except that the $u$ quantity is replaced by
$(u_2^2-u_1^2)/u$. If we take this quantity to have roughly the same
value as $u$, then our semi-quantitative conclusions are the same for
both models. Note also that this process requires one
flavour-conserving vertex. The neutral Higgs boson Yukawa interactions
feature both the $m_u$ and $m_e$ matrices for these flavour-conserving
interactions. However, with the assumed mixing pattern and because $m_u
\gg m_e$, we can approximately omit the piece proportional to the
electron mass. Similar observations regarding the Model 2 estimate
versus the Model 1 estimate, and the flavour-conserving vertex
complication, will obtain for the other processes considered below.]
The experimental bound is $\Gamma/\Gamma_{\mu} < 10^{-12}$
\cite{PDG-11}, where $\Gamma_{\mu}$ is the total width of $\mu$,
which leads to the constraint that
\begin{equation}
\left({{\epsilon m_c m_u} \over m^2_{\phi}}\right)^2 < 10^{-11},
\end{equation}
having used $u \simeq 300$ GeV. We see that typical values like
$\epsilon \simeq 0.1$ and $m_{\phi} \simeq 300$ GeV fall well within this
limit.

It is of interest to also look at $\tau$-lepton rare decays, since the
large top-quark mass contributes. The bound for $\tau^- \to \mu^- \mu^+
\mu^-$ obtained from the experimental limit $\Gamma/\Gamma_{\tau} <
10^{-5}$ \cite{PDG-11} is
\begin{equation}
\left({{\epsilon^2 m_t m_c} \over m^2_{\phi}}\right)^2 < 10^{-4},
\end{equation}
which again is easily satisfied with $\epsilon \simeq 0.1$ and $m_{\phi}
\simeq 300$ GeV. The other rare decays of the $\tau$ which can be
mediated by tree-level Higgs-boson exchange,
such as $\tau \to \mu e e$ and $\tau \to eee$,
all give much weaker constraints.

Turning now to the up-quark sector, we will examine $D^0-\bar{D}^0$
mixing and the rare decay $D^0 \to \mu^+\mu^-$. Neutral Higgs boson
exchange contributes to neutral $D$-meson mixing in both the s- and
t-channels. This leads to an extra contribution to the mass difference
given by
\begin{equation}
\Delta m = {1 \over 24}
\left({{\epsilon m_{\mu}} \over u m_{\phi}}\right)^2
m_D f^2_D \left[ {{7m^2_D} \over (m_u + m_c)^2} + 1 \right],
\end{equation}
where $m_D = 1.9$ GeV is the mass of the $D$-meson and $f_D = 200$ MeV is
its decay constant. We have used the standard vacuum-saturation
approximation to calculate this expression.
The experimental upper bound
on the mass difference is $\Delta m < 10^{-13}$ GeV \cite{PDG-11},
which leads to the limit
\begin{equation}
\left( {{\epsilon m_{\mu}} \over m_{\phi}} \right)^2 < 10^{-7}.
\end{equation}
Again, the values $\epsilon = 0.1$ and $m_{\phi} = 300$ GeV easily
satisfy this constraint.

The partial width of the decay $D^0 \to \mu^+\mu^-$ is calculated
within the vacuum saturation approximation to be
\begin{equation}
\Gamma_{\mu^+\mu^-} = {1 \over {8\pi}}
\left({{\epsilon m_{\mu} m_c} \over {u^2 m^2_{\phi}}}\right)^2
\left( {f_D \over {m_u + m_c}} \right)^2 m^5_D.
\end{equation}
The experimental bound is $\Gamma_{\mu^+\mu^-} < 10^{-17}$ GeV
\cite{PDG-11}, which leads to the constraint
\begin{equation}
\left( {{\epsilon m_{\mu} m_c} \over m^2_{\phi}} \right)^2 < 10^{-5}.
\end{equation}
Once again, this is easily satisfied with $\epsilon = 0.1$ and $m_{\phi}
= 300$ GeV.

So, we have shown that the tree-level flavour-changing neutral Higgs
boson effects within the model can easily fall within current
experimental limits on unseen decays and on neutral meson mixing. Most
spectacularly, the down-quark sector yields no bounds because the
neutrinos are constrained to be light. But, as we have just seen, the
charged-lepton and up-quark sector bounds can be met with Higgs boson
masses of the order of the electroweak scale, provided we invoke
hierarchical generation mixing. Note, however, that the mixing hierarchy
need be no more severe than what we find in the KM matrix.

\vskip 1 cm
\leftline{\it 3.3.2 Higher loop effects}
\vskip 5 mm

The main phenomenological point we want to make in this paper is that
the existence of neutral flavour-changing vertices
in quark-lepton symmetric models does not necessarily
mean that the offending electroweak Higgs bosons have to be made
artificially heavy. We feel this point is adequately demonstrated by the
bounds calculated in the preceeding subsubsection for tree-level
processes. However, there are a whole host of interesting effects that
will be induced at higher-loop levels also, and so for completeness we
include a brief discussion of some of them.

For instance, radiative decays of the second and third generation quarks
and leptons will be induced at 1-loop order. In addition to the SM graph
featuring a virtual fermion/$W$ loop, there will be new contributions
coming from charged--Higgs-boson/fermion loops and flavour-changing
neutral--Higgs-boson/fermion loops. (There are also small
1-loop contributions from a heavy-$W'$/lipton virtual pair.
See Ref.~\cite{FLV-2}
for a brief discussion.) It would be interesting to compare the
predictions for these decays in q-$\ell$ symmetric models with other
two--Higgs-doublet models \cite{higgs-12}.
For instance, in Model 1 with zero Dirac
neutrino masses there will be no contribution from either charged or
neutral Higgs bosons to $b \to s\gamma$. This amusing fact may be
important, given that the recent CLEO measurement of $B \to K^*\gamma$
is consistent with SM expectations \cite{CLEO-13}.
Similar 1-loop effects will also contribute to processes like $b \to
s\ell^+\ell^-$ where $\ell$ is a lepton.

Charged Higgs bosons will contribute to neutral-meson mixing through box
graphs, and the distinctive q-$\ell$ partnership phenomenon
in the Yukawa Lagrangians should produce interesting systematics. For
instance, in Model 2 the charged Higgs bosons have couplings
proportional to both quark and lepton mass matrices, whereas in Model 1
the coupling is exclusively through the partner lepton mass matrix. It
would be interesting to see how expectations for both of these models
compare with expectations in other two--Higgs-doublet models
\cite{higgs-12}.

Finally, we note that there will be extra contributions to the anomalous
moments of charged leptons at 1-loop level. For instance, both the
charged and neutral Higgs bosons contribute to the anomalous magnetic
moment $a_{\mu}$ of the muon. What is of some interest here is that the
vertices involved will have pieces proportional to the up-quark masses,
and so will be larger than those in the usual two--Higgs-doublet models.
Given our assumed mixing pattern, $a_{\mu}$ will be about equally
affected by $m^2_c$ and $(\epsilon m_t)^2$, which are both
about two orders of magnitude greater than the corresponding quantity in
the SM, namely $m^2_{\mu}$. The SM neutral Higgs contribution gives
roughly $a_{\mu} \simeq 10^{-13}$ \cite{Leveille-14},
so that the contribution in q-$\ell$
symmetric models is roughly $10^{-11}$. This is comfortably below the
experimental error on $a_{\mu}$ which is about $10^{-8}$.

\vskip 1 cm
\leftline{\it 3.3.3 The heavy Higgs bosons}
\vskip 5 mm

We end this section with a few words about the phenomenology of the
heavy Higgs bosons in Higgs sector B, namely the neutral field $H$
and the charged and coloured field $\chi_1$.

Particle $H$ is expected to
be heavy because it is associated with leptonic colour and discrete
symmetry breaking. However, it couples only very weakly to standard
particles and is thus very difficult to produce in the laboratory. Its
only direct Yukawa couplings are to liptons, not leptons, and it does not
couple to any of the electroweak gauge bosons. Its tree-level couplings
to standard particles occur therefore only through mixing with
$h'_{1,2}$, which we expect to be quite suppressed. We thus expect the
actual phenomenological bound on $H$ to be very weak, although the
generic expectation is that it ought to be relatively heavy.

The coloured scalar $\chi_1$ contributes to neutral meson mixing, as
discussed in Ref.~\cite{DH-15}. The generic bound obtained is that
$h_2/m_{\chi} < 10^{-3}$GeV$^{-1}$, where $h_2$ is relevant Yukawa
coupling constant. We note that this is a weak constraint.

\vskip 1 cm
\leftline{\large \bf 4. Conclusion}
\vskip 5 mm

The simplest realistic Higgs sector in models with leptonic colour and a
quark-lepton discrete symmetry contains two electroweak Higgs doublets.
We have demonstrated that the effective two--Higgs-doublet model
obtained at the electroweak scale is unusual and interesting, because
Yukawa coupling constants are in important instances
proportional to the mass
matrix of the discrete symmetry partner of the fermion in question,
rather than of the fermion itself. In particular, the flavour-changing
neutral vertices are always proportional to the mass of the partner
fermion. If neutrinos are Dirac particles then this implies that
flavour-changing neutral Higgs effects in the down-quark sector are
negligible, and so all of the traditional constraints like those from
neutral kaon mixing are absent. The largest flavour-violating effects
occur in the charged-lepton and up-quark sectors, but we showed that
these effects are typically undetectable if a Kobayashi-Maskawa pattern
of inter-generation mixing is invoked.
We conclude, therefore, that electroweak
Higgs bosons with masses in the $100$-GeV range are perfectly
acceptable, even though some of them mediate neutral flavour-changing
processes at tree-level.

\vskip 1 cm
\centerline{\large \bf Acknowledgements}
\vskip 5 mm

YL wishes to thank Xiao-Gang He and Andrew J. Davies for very useful
discussions. RRV would like to thank Henry Lew
and Robert Foot for interesting communications and Xiao-Gang He for
useful discussions. He is supported by grants from the University of
Melbourne.


\begin{thebibliography}{99}

\bibitem{FL-1} R. Foot and H. Lew, Phys.\ Rev.\ D{\bf 41}, 3502 (1990).

\bibitem{FLV-2} R. Foot, H. Lew and R. R. Volkas, Phys.\ Rev.\ D{\bf
44}, 1531 (1991).

\bibitem{FLV-3} R. Foot, H. Lew and R. R. Volkas, Phys.\ Rev.\ D{\bf
44}, 859 (1991) and (E) {\it ibid.} D{\bf 47}, 1272 (1993);
G. C. Joshi and R. R. Volkas, Phys.\ Rev.\ D{\bf 45}, 1711 (1992).

\bibitem{FL-4} R. Foot and H. Lew, Mod.\ Phys.\ Lett.\ A{\bf 7}, 301
(1992).

\bibitem{LV-5} H. Lew and R. R. Volkas, Phys.\ Rev.\ D{\bf 47}, 1356
(1993).

\bibitem{FLV-6} R. Foot, H. Lew and R. R. Volkas, Int.\ J. Mod.\ Phys.\
A{\bf 8}, 983 (1993).

\bibitem{glueballs-7} For studies of glueballs in a
related model see E. D. Carlson et al., Phys.\ Rev.\ D{\bf 44}, 1555
(1991) and R. Foot and H. Lew, Purdue University
Report No.\ PURD-TH-92-3 (unpublished).

\bibitem{seesaw-8} M. Gell-Mann, P. Ramond and R. Slansky, in {\it
Supergravity}, eds.\ P. Van Nieuwenhuizen and D. Z. Freedman
(North-Holland, Amsterdam, 1979); T. Yanagida in {\it Proc. Workshop on
Unified Theory and Baryon Number of the Universe}, eds. A. Sawada and H.
Sugawara (KEK, Tsukuba-Gun, Ibaraki-Ken, Japan, 1979); R. N. Mohapatra
and G. Senjanovic, Phys.\ Rev.\ Lett.\ {\bf 44}, 912 (1980).

\bibitem{FLV-9} R. Foot, H. Lew and R. R. Volkas, University of
Melbourne Report No.\ UM-P-91/46 (Mod.\ Phys.\ Lett., in press).

\bibitem{FL-10} R. Foot and H. Lew, Nuovo Cim.\ A{\bf 104}, 167 (1991).

\bibitem{PDG-11} Particle Data Group, Phys.\ Rev.\ D{\bf 45}, S1
(1992).

\bibitem{higgs-12} For a review see J. F. Gunion, H. E. Haber, G. Kane
and S. Dawson, {\it Higgs Hunter's Guide} (Addison-Wesley, 1990, Redwood
City).

\bibitem{CLEO-13} CLEO Collaboration, forthcoming report.

\bibitem{Leveille-14} J. P. Leveille, Nucl.\ Phys.\ B{\bf 137}, 63
(1978).

\bibitem{DH-15} A. J. Davies and X.-G. He, Phys.\ Rev.\ D{\bf 43}, 225
(1991).

\end{thebibliography}
\end{document}